\documentclass[journal=jacsat,manuscript=article]{achemso}

\usepackage[version=3]{mhchem} % Formula subscripts using \ce{}

\author{Kamal Choudhary (0000-0001-9737-8074)}
\email{kamal.choudhary@nist.gov}
\affiliation[National Institute of Standards and Technology]
{Material Measurement Laboratory, National Institute of Standards and Technology, Gaithersburg, 20899, MD, USA.}
\alsoaffiliation[Theiss Research]
{Theiss Research, La Jolla, 92037, CA, USA.}

 \author{Brian DeCost (0000-0002-3459-5888)}
\affiliation[National Institute of Standards and Technology]
{Materials Measurement Laboratory, National Institute of Standards and Technology, Gaithersburg, 20899, MD, USA.}

\author{Lily Major (0000-0002-5783-8432)}
\affiliation[Aber]
{Department of Computer Science, Aberystwyth University, SY23 3DB, UK}
\alsoaffiliation[Rutherford Appleton Laboratory]
{Scientific Computing Department, Rutherford Appleton Laboratory, Science and Technology Facilities Council, Harwell Campus, Didcot, OX11 0QX, UK}

\author{Keith Butler (0000-0001-5432-5597)}
\affiliation[Rutherford Appleton Laboratory]
{Scientific Computing Department, Rutherford Appleton Laboratory, Science and Technology Facilities Council, Harwell Campus, Didcot, OX11 0QX, UK}

\author{Jeyan Thiyagalingam (0000-0002-2167-1343)}
\affiliation[Rutherford Appleton Laboratory]
{Scientific Computing Department, Rutherford Appleton Laboratory, Science and Technology Facilities Council, Harwell Campus, Didcot, OX11 0QX, UK}

\author{Francesca Tavazza (0000-0002-5602-180X)}
\affiliation[National Institute of Standards and Technology]
{Materials Measurement Laboratory, National Institute of Standards and Technology, Gaithersburg, 20899, MD, USA.}
\title{Unified Graph Neural Network Force-field for the Periodic Table}

\abbreviations{IR,NMR,UV}

\begin{document}

\begin{abstract}

Classical force fields (FF) based on machine learning (ML) methods show great potential for large scale simulations of materials. MLFFs have hitherto largely been designed and fitted for specific systems and are not usually transferable to chemistries beyond the specific training set. We develop a unified atomisitic line graph neural network-based FF (ALIGNN-FF) that can model both structurally and chemically diverse materials with any combination of 89 elements from the periodic table. To train the ALIGNN-FF model, we use the JARVIS-DFT dataset which contains around 75000 materials and 4 million energy-force entries, out of which 307113 are used in the training. We demonstrate the applicability of this method for fast optimization of atomic structures in the crystallography open database and by predicting accurate crystal structures using genetic algorithm for alloys. 
\end{abstract}

%%%%%%%%%%%%%%%%%%%%%%%%%%%%%%%%%%%%%%%%%%%%%%%%%%%%%%%%%%%%%%%%%%%%%
%% Start the main part of the manuscript here.
%%%%%%%%%%%%%%%%%%%%%%%%%%%%%%%%%%%%%%%%%%%%%%%%%%%%%%%%%%%%%%%%%%%%%
\section{Introduction}

% - Atomist. sim. large scale req. FF
% - Hard to develop FF for generalized chemical species, multi-component
% - MLFF review
% - Search for materials with GA requires structure optimization, and is the bottleneck when done with DFT
% - This particular FF will be useful

% -Only lattice constants as an error bar
% - Validate on 5 gdr/rdf final rdf
% - Manually altered POSCAR
% - Different spacegroup (brutal test), out of cubics
% - Keep Fig 3
% - Add GA

% [Random structure searching (maximally unbiased), GA, minima hoping, USPEX] review on Xtal search prediction

Large scale atomistic simulation of multi-component systems is a difficult task but they are highly valuable for industrial applications such as designing alloys, designing electrical contacts, touch screens, transistors, batteries, composites and catalysts \cite{ogale2006thin,andersson2006toward,liang2013classical,li2009transfer}. Quantum chemistry methods, such as density functional theory, are obvious approaches to simulate such systems, however, they are computationally very expensive for large systems \cite{srolovitz2012atomistic}. Classical force-fields, or interatomic potentials, such as embedded-atom method (EAM), modified embedded-atom method (MEAM), reactive bond-order (ReaxFF), charge-optimized many-body (COMB) etc., \cite{daw1984embedded,liang2012variable,brenner1990empirical,van2001reaxff,brenner2000art} can be used for such simulations but they are usually parameterized for a very narrow chemical phase-space limiting their applicability and transferability. Moreover, it can be quite strenuous and time consuming to develop such traditional classical FFs.

%To quote noble laureate Herbert Kroemer: “The interface is the device”.
Recently machine learning based FFs \cite{poltavsky2021machine,choudhary2022recent} are being used to systematically improve accuracy of FFs and have successfully been used for multiple systems. One of the pioneer MLFFs were developed by Behler-Parinello in 2007 using neural network \cite{behler2007generalized}. It was initially used for molecular systems and now has been extended to numerous other applications \cite{unke2021machine}. Although neural network is one of the most popular regressors, other methods such as Gaussian processes-based Gaussian
approximation potentials (GAP) \cite{bartok2010gaussian}, as well as linear regression and basis functions-based spectral neighbor analysis potential (SNAP) \cite{wood2018extending} have also been thoroughly used. Such FFs use two and three body descriptors to describe local environment. Other popular MLFF formalisms include smooth overlap of atomic positions (SOAP) \cite{bartok2013representing}, moment tensor potentials (MTP) \cite{shapeev2016moment,novikov2020mlip}, symbolic regression \cite{hernandez2019fast} and polynomial-based approaches \cite{drautz2019atomic}. One of the critical issues in developing and maintaining classical force-fields is that they are hard to update with software and hardware changes. Luckily, MLFFs are more transparently developed and maintained than other classical FFs. A review article on this topic can be found elsewhere \cite{unke2021machine}. Nevertheless, early-generation MLFFS are also limited to a narrow chemical space and may require hand-crafted descriptors which may take time to be identified.

Graph neural network (GNN) based methods have shown remarkable improvements over descriptor based machine learning methods and can capture highly non-Euclidean chemical space \cite{schutt2018schnet,xie2018crystal,chen2019graph,chen2022universal,kearnes2016molecular,choudhary2022recent,gilmer2017neural,klicpera2020fast,batzner20223}. GNN FFs have been also recently proposed and are still in the development phase \cite{park2021accurate,chmiela2018towards}. We developed an atomistic line graph neural network (ALIGNN) in our previous work\cite{choudhary2021atomistic} which can capture many body interactions in graph and successfully models more than 70 properties of materials, either  scalar or vector quantities, such as formation energy, bandgap, elastic modulus, superconducting properties, adsorption isotherm, electron and density of states etc \cite{choudhary2021atomistic,choudhary2022graph,choudhary2022designing,choudhary2022deep,kaundinya2022prediction,gurunathan2022rapid}. The same automatic differentiation capability that allows training these complex models allows for physically consistent prediction of quantities that must follow conservation laws, such as forces and energies; this enables GNNs to be used in molecular dynamic simulations to quickly identify relaxed or equilibrium states of complex systems. However, there is a need for a large and diverse amount of data to train unified force-fields. 

In this work, we present a dataset of energy and forces with 4 million entries for around 75000 materials in the JARVIS-DFT dataset which have been developed over past 5 years \cite{choudhary2020joint}. We extend the ALIGNN model to also predict derivatives that are necessary for FF formalism. There can be numerous applications of such a unified FF, however, in this work we limit ourselves to pre-optimization of structures, genetic algorithm based structure, and molecular dynamics applications. The developed model will be publicly available on the ALIGNN GitHub page (\url{https://github.com/usnistgov/alignn}) with several examples and a brief documentation. 

% This project is a part of the diverse NIST-JARVIS infrastructure (\url{https://jarvis.nist.gov/}) \cite{choudhary2020joint}.

%\section{Results and discussion}

\section{Methodology}
A flow-chart for training ALIGNN-FF is shown in Fig. 1a. To train the FF, we use a large DFT dataset, JARVIS-DFT, which contains about 75000 materials with a variety of atomic structures and chemistry and has been generated over the last 5 years. JARVIS-DFT is primarily based on Vienna Ab initio Simulation Package (VASP) \cite{kresse1996efficient,kresse1996efficiency} software and OptB88vdW \cite{klimevs2009chemical} functional but also contains data obtained using other functionals and methods. In this work, only OptB88vdW-based data has been used. The OptB88vdW functional was shown to be very well applicable to solids in Ref. \cite{klimevs2009chemical} and, ever since, it has been used to model rare-gas dimers and metallic, ionic, and covalent bonded solids, polymers, and small molecular systems \cite{choudhary2018elastic}. Energies and forces are available for each structure optimization and elastic constant calculation runs. The total number of such entries is around 4 million. Although it would be justified to train on the entire dataset, we choose to use only a subset of it because of the computational budget and hardware requirements available to us. Instead of the 4 million datapoints, we use 307113 points i.e. more than an order magnitude less, by taking unique set of first, last, middle, maximum energy and minimum energy structures only. If some snapshots for a run are identical, say the last step and the minimum energy configuration are the same, then we only include one of them. The dataset consists of perfect structures only.
%i.e. without any defects and energy-volume distortions.

%2764017 stress, mad 12.749869945020103, (32327.24990773, -9052.16582282 kbar)
 We convert the atomic structures to a graph representation using atomistic line graph neural network (ALIGNN). Details on ALIGNN can be found in the related paper \cite{choudhary2021atomistic}. In brief, each node in the atomistic graph is assigned 9 input node features based on its atomic species: electronegativity, group number, covalent radius, valence electrons, first ionization energy, electron affinity, block and atomic volume. The inter-atomic bond distances are used as edge features with radial basis function up to 8 $\textrm{\AA}$ cut-off. We use a periodic 12-nearest-neighbor graph construction. This atomistic graph is then used for constructing the corresponding line graph using interatomic bond-distances as nodes and bond-angles as edge features. ALIGNN uses edge-gated graph convolution for updating nodes as well as edge features. One ALIGNN layer composes an edge-gated graph convolution on the bond graph with an edge-gated graph convolution on the line graph. The line graph convolution produces bond messages that are propagated to the atomistic graph, which further updates the bond features in combination with atom features. 

In this work, we developed the functionality for atomwise and gradient predictions. 
Quantities related to gradients of the predicted energy, such as forces on each atom, are computed by applying the chain rule through the automatic differentiation system used to train the GNN, thus preserving the law of conservation of energy.
 In a closed system, the forces on each atom 
$i$ depend on its position with respect to every other particle $j$ through a force-field as:
%https://klyshko.github.io/teaching/2019-03-01-teaching
\begin{equation} 
m_i\frac{d^2r_i(t)}{dt^2}=\sum_{j}F_{ij}(t)=-\sum_{j}\nabla_iU(r_{ij}(t))
\end{equation} 
 where $r_{ij}$ is the distance between atom $i$ and $j$.
 For stress predictions, we use the Virial stress \cite{subramaniyan2008continuum} which is a measure of mechanical stress on an atomic scale for homogeneous systems:
%https://www.google.com/search?q=virial+stress&rlz=1C1GCEV_en&sxsrf=ALiCzsYghr2GvmNL8I6yqVPewG9PUy3l7Q:1657574815077&source=lnms&tbm=isch&sa=X&ved=2ahUKEwie96mw4_H4AhXlFFkFHUWqCSMQ_AUoAXoECAIQAw&biw=1182&bih=637&dpr=1.63&safe=active&ssui=on#imgrc=bndGox8sMioE0M
 
 \begin{equation} 
\sigma_{\alpha\beta}=-\frac{1}{V}\sum_{i}\sum_{i\neq j}F_{ij}^{\alpha}r_{ij}^{\beta}+m_i v_i^{\alpha} v_i^{\beta}
\end{equation}

The 307113 data points are split into 90:5:5 ratio for training, validation and testing. We train the model for 250 epochs using the same hyper-parameters as in the original ALIGNN model \cite{choudhary2021atomistic}. ALIGNN is based on deep graph library (DGL) \cite{wang2019deep}, PyTorch \cite{paszke2017automatic} and JARVIS-Tools packages \cite{choudhary2020joint}. We optimize a composite loss function ($l$) with weighted mean absolute error terms for both forces and energies:

% \begin{equation}
% l = \frac{1}{N_{atoms}} \left|E^{DFT} - E^{GNN}\right| + \frac{w}{3N_{atoms}} \sum_i^{N_{atoms}} \left\|F_i^{DFT} - F_i^{GNN}\right\|
% \end{equation}

\begin{equation}
l =\left|E^{DFT} - E^{GNN}\right| + w \sum_i^{N_{atoms}} |F_i^{DFT} - F_i^{GNN}|
\end{equation}

where, $E^{DFT}$ and $E^{GNN}$ are energies per atom using DFT and ALIGNN, $F_i^{DFT}$ and $F_i^{GNN}$ are forces acting on an atom using DFT and ALIGNN,  $w$ is a tunable  weighting factor scaling the force contribution to the loss relative to the energies; in this work we set $w = 10$.
Without such weighting, it is difficult to learn forces as they vary within a wide range. 

The ALIGNN-FF model has been integrated with atomic simulation environment (ASE) \cite{larsen2017atomic} as an energy, force and stress calculator for structure optimization and MD simulation. This calculator can be used for optimizing atomic structures, using genetic algorithm \cite{johannesson2002combined}, and running molecular dynamics simulations, for example  constant-temperature, constant-volume ensemble (NVT) simulations. The structural relaxations are carried out with the fast inertial relaxation engine (FIRE) \cite{bitzek2006structural}, available in ASE.
In order to predict equation of state/ energy-volume-curve (EV) simulation, we apply volumetric strains between ranges of -0.1 to 0.1 with an interval of 0.01.

% The defect structures are created by deleting a unique Wyckoff site atom in a cell of at least 1.5 nm size in each direction. The vacancy formation energies are calculated using the formula: \begin{math}E_{vacancy}=E_{defect}-E_{perfect}+\mu\end{math}
% where, $E_{vacancy}$ is the vacancy formation energy, $E_{defect}$ is the energy of the defect structure with an atom missing, $E_{perfect}$ is the energy of the perfect structure, $\mu$ is the chemical potential used as energy per atom of the most stable structure of an element. 

% The surface energies are calculated as: \begin{math}E_{surface}=(E_{slab}-N_{slab}*E_{bulk-patom})/2A\end{math}
% where, $E_{surface}$ is the surface energy,$E_{slab}$ is the total energy of slab with at least 2.5 nm thickness and 1.8 nm vacuum padding, $N_{slab}$ is the number of atoms in the slab, $E_{bulk-patom}$ is the energy per atom of the bulk relaxed counterpart of the slab, $A$ is the area of the surface. All slabs were created with vacuum along the z-direction. 

The interfaces are generated using Zur algorithm \cite{zur1984lattice,mathew2016mpinterfaces,ding2016computational,choudhary2020efficient} as implemented in the JARVIS-Tools. For a few cases, we compare the ALIGNN-FF data with classical-forcefield embedded atom method (EAM) \cite{daw1983semiempirical,pun2015interatomic,farkas2020model} and density functional theory based GPAW packages \cite{enkovaara2010electronic}.

%The work of adhesion is calculated as: \begin{math}W_{adhesion}=(E_{film}+E_{substrate}-E_{interface})/A\end{math} where, $W_{adhesion}$ is the work of adhesion, $E_{film}$ total energy of the top film slab, $E_{substrate}$ total energy of the substrate slab, $E_{interface}$ is the total energy of the interface by putting the two slabs together.

\section{Results and discussion}

\subsection{Performance on test set}

 \begin{figure}[hbt!]
    \centering
    \includegraphics[trim={0. 0cm 0 0cm},clip,width=0.95\textwidth]{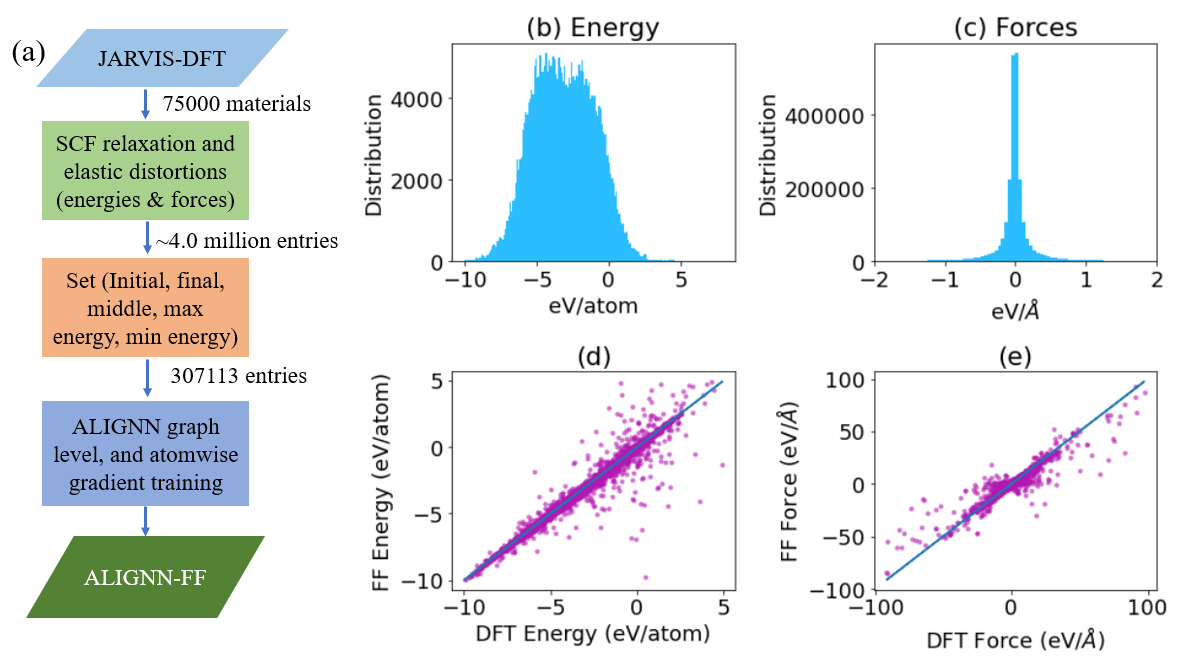}
    \caption{Schematic showing flow chart for developing the force-field, visualization of available data and model performance on test dataset. The left panel shows how dataset was obtained from JARVIS-DFT and model was trained. Fig. a and b shows energy and force-distribution in the dataset. While energy dataset ranges from -9.98 to 9.97 eV and is well-dispersed over this range, the force dataset varies from -591.12 to 591.40 eV/$\textrm{\AA}$ and is highly localized around zero. The force-dataset contains the x,y,z force on each atom. Fig. d and Fig. e shows energy and force predictions on the test set.}
\end{figure}
We find that the energy per atom dataset (with 307113 entries) varies from -9.98 eV/atom to 9.97 eV/atom. The force dataset contains 9593385 entries (resulting from x,y,z forces on each atom) is mostly centered around zero, but ranges from -591.12 eV/$\textrm{\AA}$ to 591.40 eV/$\textrm{\AA}$. The mean absolute deviation (MAD) for energies and forces are 1.80 eV and 0.10 eV/$\textrm{\AA}$ respectively. The distributions of the energy and force used in the training are shown in Fig. 1b and 1c, respectively. The total dataset was split in 90:5:5 train-validation-test sets. We show the performance on test set for energies and forces in Fig. 1d and 1e respectively. We note that these points represent a variety of chemistry and structures unlike usual MLFFs which are usually focused on a specific chemistry. The mean absolute error (MAE) of the predicted energy per atom and force component per atom are 0.086 eV and 0.047 eV/$\textrm{\AA}$ respectively. In comparison to the previous ALIGNN model for energy only, which has an MAE of 0.03 eV/atom, 0.086 eV/atom might seem high, however, we note that the previous energy model was trained on relaxed structures only while the current model also captures several un-relaxed structures which can be much higher in energy scale. The MAD:MAE ratios for energies and forces are 20.93 and 2.12 which are high. 

While the above results are for force tunable weighting factor of 10, we also train the models with other weigthing factors as shown in Table 1. We find that as we increase the weighting factors, the MAEs increase for energies increase but decrease for forces. As the MAD for forces is 0.1 eV/$\textrm{\AA}$, we choose to work with the model obtained for the lowest MAE for forces and to analyze its applications in the rest of the paper. Nevertheless, we share model parameters for other weighting factors for those interested in analyzing its effect on property predictions.

\begin{table}
\caption{Effect of different weighting factors for energy and force predictions.}
% python l.py >out
\begin{tabular}{@{}lll@{}}
   \hline
Weight & MAE-Energies (eV/atom) & MAE-Forces (eV/$\AA$)\\
   \hline
0.1&0.034&0.092\\
0.5&0.044&0.089\\
1.0&0.051&0.088\\
5.0&0.082&0.054\\
10.0&0.086&0.047\\
   \hline

\end{tabular}
\end{table}

% The results also suggest it is easier to learn energy data than forces.
%alloy choudhuri2019enhancing
\subsection{Energy-volume curves}
The energy-volume (E-V) curves are crucial to understand the behavior of an FF. We obtain EV-curves for a few test case materials by applying volumetric strains between ranges of -0.1 to 0.1 with an interval of 0.01. Specifically, we compare the energy volume curves for Ni$_3$Al (JVASP-14971), Al$_2$CoNi (JVASP-108163), CrFeCoNi (4 atoms cell with spacegroup number 216 and lattice parameter of 4.007 $\textrm{\AA}$), NaCl (JVASP-23862), MgO (JVASP-116), and BaTiO$_3$ (JVASP-8029) using EAM potential \cite{pun2015interatomic,farkas2020model}, GPAW DFT, and ALIGNN-FF in Fig. 2. The energy scales for these methods differ so we align them with respect to the corresponding minimum energies, for comparison. We used less EV-curve points for a structure in GPAW to save computational cost. We notice all the EV-curves are parabolic in nature and smooth, indicating a smooth potential energy surface. 

We find that the EV curves from these methods coincide near the minimum for all systems but for Al$_2$CoNi and CrFeCoNi, the GPAW equilibrium volume is slightly smaller than for EAM and ALIGNN-FF, suggesting that the lattice constants for ALIGNN-FF and EAM might be overestimated compared to GPAW. Nevertheless, EAM and ALIGNN-FF data agree well. Comparing EAM and ALIGN-FF, it's important to remember that GNNs have no fundamental limitation to number of species they can model (i.e. high chemical diversity), and can in principle even extrapolate to species not contained in the training set, which is extremely powerful compared to conventional FFs like EAM.

% While, EAM and GPAW curves are very similar, the shapes of the ALIGNN-FF EV-curves may also differ compared to GPAW. For instance, in d,e,f example, we find that for contraction region (i.e. when atoms are closer), the two EV-curves differ. Nevertheless, this behavior is not universal. We note that we do not use any energy-volume curve data in the training of ALIGNN-FF. Also, there are no data available in FeCoNiCr in the training data suggesting that ALIGNN-FF might be working well for extrapolation and doesn't work as a look-up table only. This behavior suggests a strong transferability feature of ALIGNN-FF.

Note that there are many other conventional FF repositories available (such as Inter-atomic Potential Repository \cite{becker2013considerations} and JARVIS-FF \cite{choudhary2017evaluation}) which contain data for a variety of systems. It is beyond the scope of the current work to compare all of them with ALIGNN-FF, however, it would be an interesting effort for the future work. 

 \begin{figure}[hbt!]
    \centering
    \includegraphics[trim={0. 0cm 0 0cm},clip,width=1.0\textwidth]{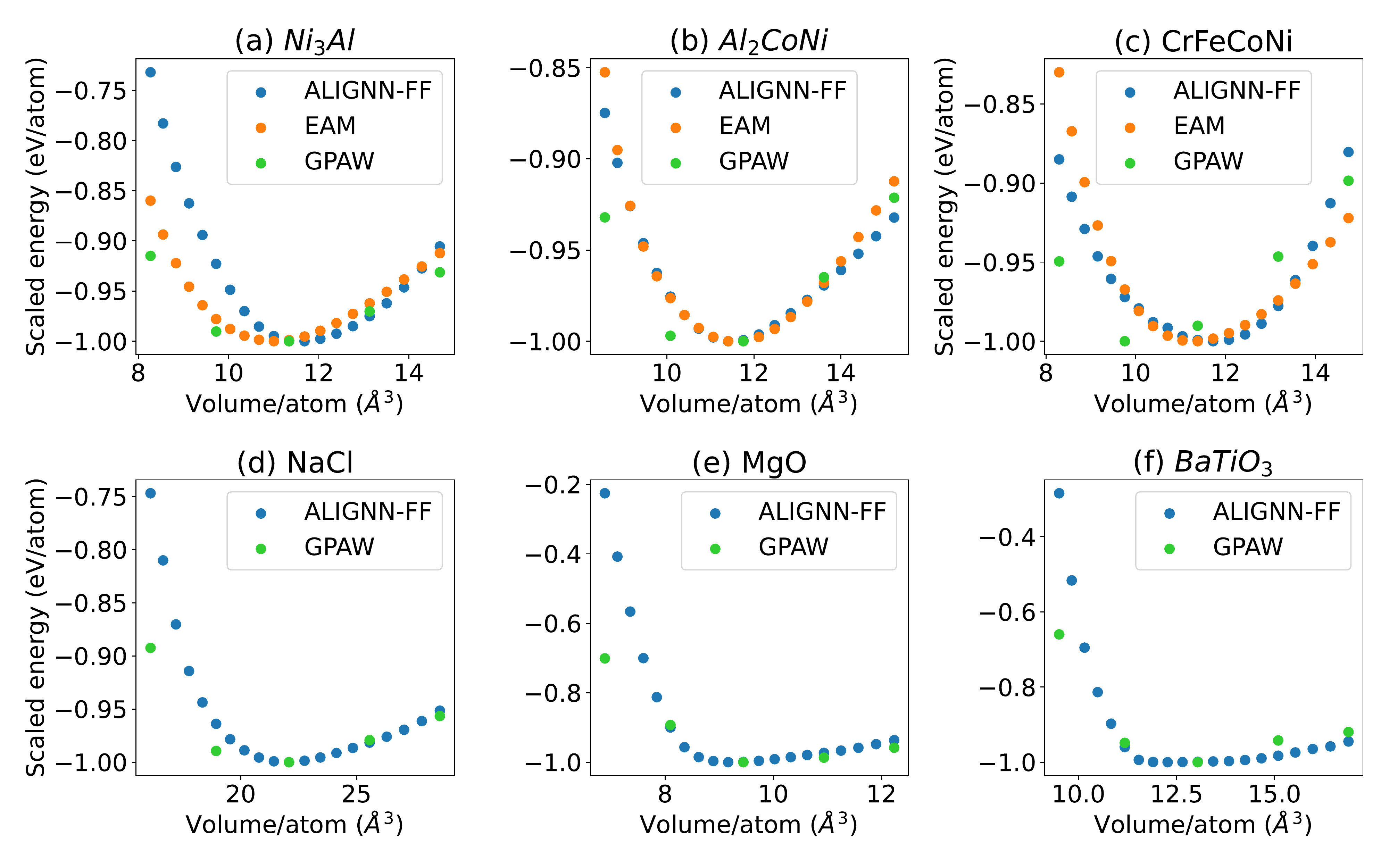}
    \caption{Energy-volume/expansion-contraction curves for a few example systems: a) Ni$_3$Al, b) Al$_2$CoNi, c) CrFeCoNi, d) NaCl, e) MgO, f) BaTiO$_3$ with ALIGNN-FF, EAM force-fields and GPAW DFT methods. We apply volumetric strains between ranges of -0.1 to 0.1 with an interval of 0.01. For GPAW, we use an interval of 0.05.}
    \label{fig:ev-curve}
\end{figure}

% [ADD Si,SiO2, CeO2]

% [TODO: 1. Add Quinary , 2. Add DFT points, 3. Structure where we can beat EAM]
While the above examples are for  individual crystals, it is important to distinguish different polymorphs of a composition system for materials simulation (i.e. structural diversity). In Fig. 3, we analyze the energy-volume (EV) curve of four systems and their polymorphs using ALIGNN-FF. We choose such four example systems because representative of different stable structures. In general, however, the EV-curve can be computed for any arbitrary system and structure. In Fig. 3a, we show the EV-curve for 4 silicon materials (JARVIS-IDs: JVASP-1002, JVASP-91933, JVASP-25369, JVASP-25368) with diamond cubic correctly being the lowest in energy. Similarly, the EV-curve for naturally prevalent SiO$_2$ systems (JARVIS-IDs:JVASP-58349, JVASP-34674, JVASP-34656, JVASP-58394), binary alloy Ni$_3$Al (JARVIS-IDs:JVASP-14971, JVASP-99749, JVASP-11979) and vdW bonded material MoS$_2$ (JARVIS-IDs:JVASP-28733, JVASP-28413, JVASP-58505) all have the correct structure corresponding to the minimum energy.  Therefore, while the MAE for our overall energy model is high, such model is able to distinguish polymorphs of compounds with meV level accuracy which is critical for atomistic applications.

 \begin{figure}[hbt!]
    \centering
    \includegraphics[trim={0. 0cm 0 0cm},clip,width=0.95\textwidth]{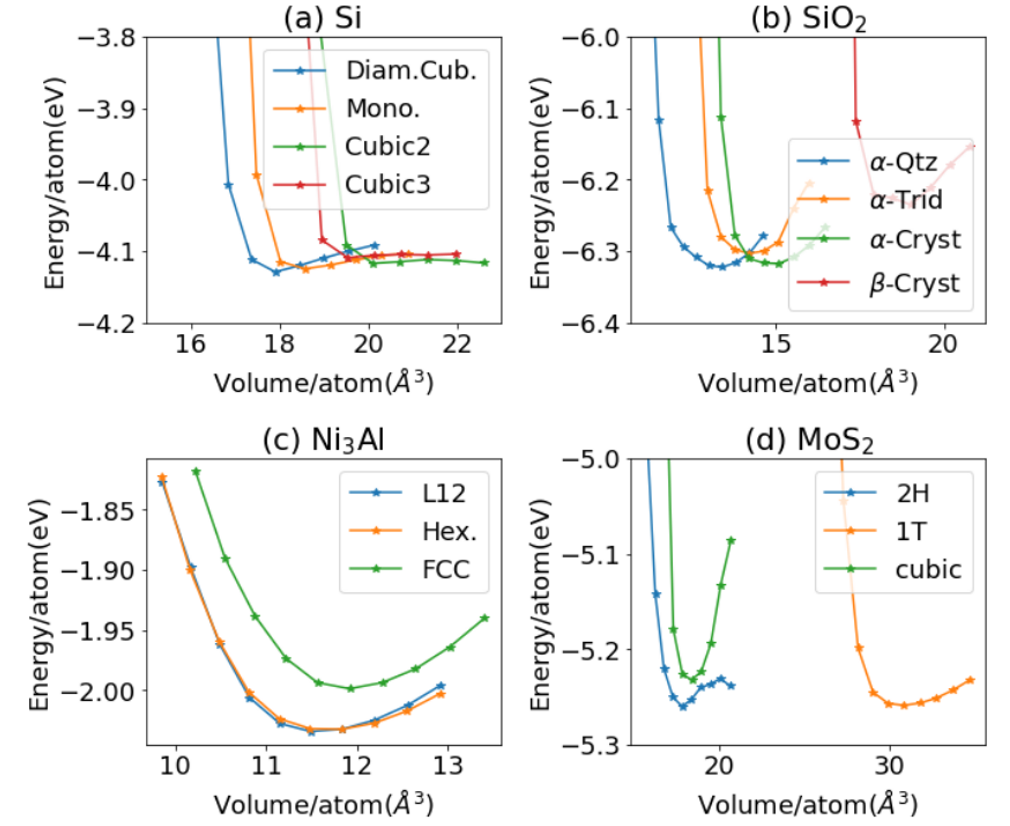}
    \caption{Energy-volume/expansion-contraction curves for a few example systems: a) silicon, b) SiO$_2$, c) Ni$_3$Al, d) MoS$_2$ polymorphs. We optimize the structures and then apply volumetric strains between ranges of -0.05 to 0.05 with an interval of 0.01.}
\end{figure}

\subsection{Lattice constants and formation energies}
%{latt_form.png}
 \begin{figure}[hbt!]
    \centering
    \includegraphics[trim={0. 0cm 0 0cm},clip,width=0.95\textwidth]{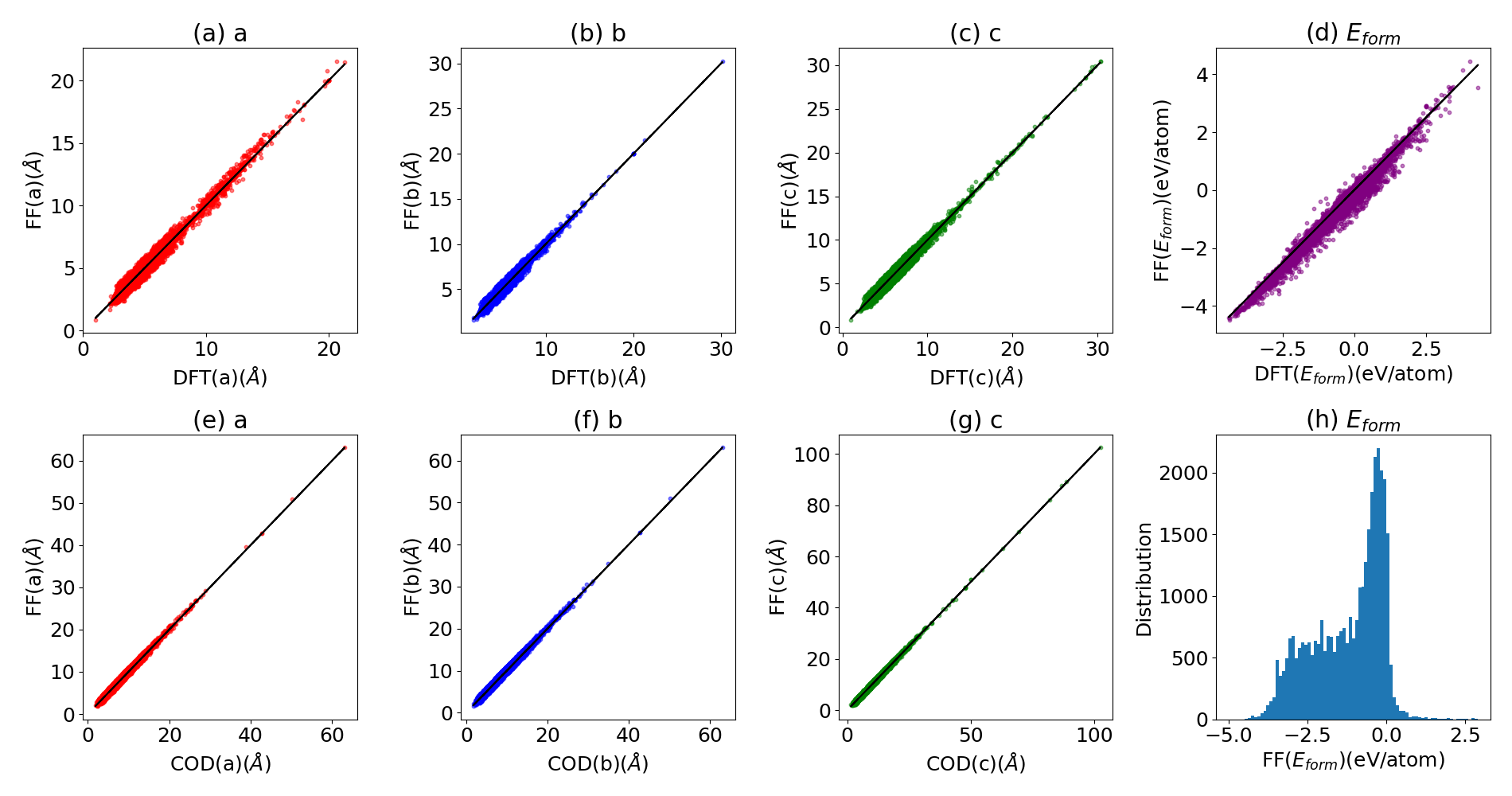}
    \caption{Comparison of DFT and FF data for a) lattice constant in x-direction, b) lattice-constant in y direction, c) lattice constant in z-direction, d) formation energies for stable binary solids in JARVIS-DFT. Comparison of crystallography open database (COD) and FF lattice constants in e) x-direction, f) y-direction, g) z-direction and h) formation energy distribution.}
\end{figure}

In this section, we compare the lattice constants and formation energies for elemental and multi-component solids from the JARVIS-DFT after optimizing them using ALIGNN-FF as shown in Fig. 4a-4d. For ALIGNN-FF we choose solids with less than 10 atoms in a cell from the JARVIS-DFT database. We optimize 23495 such materials, and present the results in Fig 4a-4d. The formation energies require chemical potential of elemental systems. The energy per atom of elemental solid systems with minimum energy are used as chemical potentials. We optimize the lattices with the FIRE algorithm as implemented in ASE and find reasonable agreement between the DFT and ALIGNN-FF lattice constants. The MAE for a,b,c lattice constants are 0.11, 0.11 and 0.13 $\textrm{\AA}$ respectively.The mean absolute error (MAE) for formation energies between ALIGNN-FF and JARVIS-DFT is 0.08 eV/atom which is reasonable for pre-screening applications. Note, the above validation is different from the performance measurement in Fig. 1 for the 5 \% test dataset because we optimize the crystals rather than directly predicting the formation energies on unrelaxed structures. Similarly, we apply the ALIGNN-FF on the crystallography open database (COD). The COD database contains more than 431778 atomic structures with different types of chemical bondings and environments. We optimize 34615 structures in the COD database with number of atoms in a cell less than 50 and show the results in Fig. 4e-4h. Here, we find the MAE for a, b, c lattice parameter as 0.20, 0.20 and 0.23 $\textrm{\AA}$ respectively. Most of the systems in COD have been derived from experiments, and hence we see many of them have negative formation  energies as shown in Fig. 4h.

\subsection{Genetic algorithm based structure search}

Computational prediction of the ground-state structures of a chemical system is a challenging task. Some of the common methods for this task include genetic algorithm (GA), simulated annealing and basin or minima hopping \cite{ji2010comparing}. In the following examples, we show using GA together with ALIGNN-FF to search for crystal structures of Ni-Al and Cu-Al example systems. Genetic algorithms mimic the biological evolution process to solve optimization problems. A GA optimization consists of  1) inheritance, 2) mutation, 3) selection, and 4) crossover operations to produce new structures and search for better survivors from generation to generation based on "survival of the fittest" idea which in our case would be energetics criteria. While all GAs follow similar strategy, the details of the individual operations can vary a lot from problem to problem and can be critical to search efficiency. 
We start with face-centered cubic (FCC) structures of the individual components (FCC Al (JVASP-816) and FCC Ni (JVASP-943) for Ni-Al search and FCC Al and FCC Cu (JVASP-867) for Cu-Al search) as the initial population. We perform relaxation of these systems with ALIGNN-FF. We choose a population size of 10 individuals and create the initial population by randomly selecting the elements. We use 40 generations to evolve the system and store the entries which are also relaxed with ALIGNN-FF.

After this example GA search for structures, we plot the convex hull diagram of these systems in Fig. 5. We find that the GA predicts AB and A$_3$B compounds which are in fact observed experimentally in such binary alloys \cite{johannesson2002combined,van2002automating}. Additionally, Ni$_3$Al (spacegroup: Pm-3m) is known to be one of the best performing super-alloys \cite{johannesson2002combined} which is reproduced in the above example. We also found that the formation energy of this structure (-0.47 eV/atom) is similar to Johannesson's\cite{johannesson2002combined} findings of -0.49 eV/atom.Although the above example is carried out for binary systems,in principle the same methodology  can be applied for any other systems as well.

 \begin{figure}[hbt!]
    \centering
    \includegraphics[trim={0. 0cm 0 0cm},clip,width=0.95\textwidth]{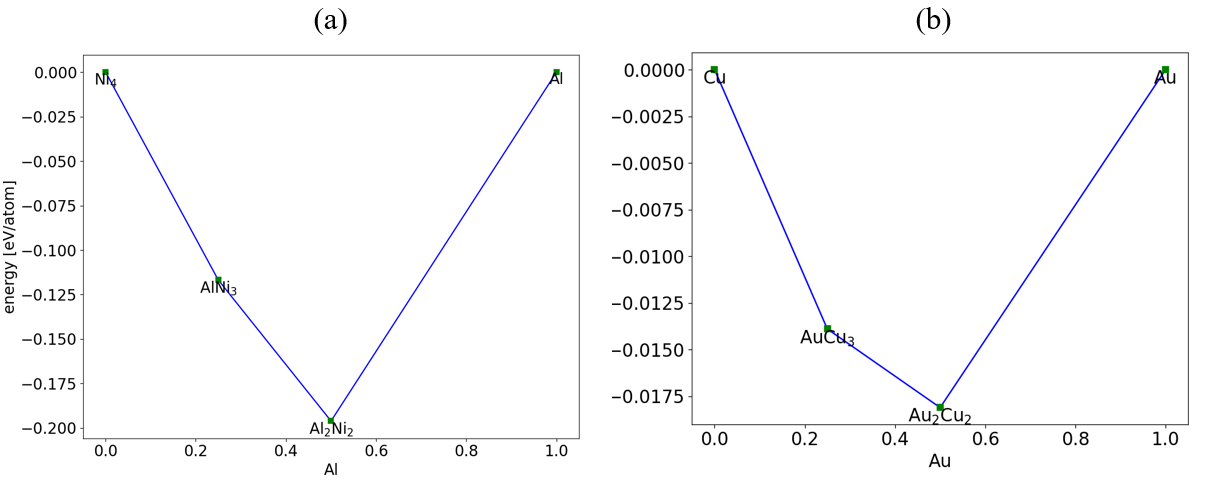}
    \caption{Convex hulls obtained from a genetic algorithm search for alloys starting with elemental solids only. The x-axis values represent mole fraction of corresponding element. We show the stable structures only for a) Ni-Al system, b) Cu-Au system.}
\end{figure}

\subsection{MD-NVT simulations}

% .[TODO: binary, quinary PD, a fe DFT points]
 \begin{figure}[hbt!]
    \centering
    \includegraphics[trim={0. 0cm 0 0cm},clip,width=0.95\textwidth]{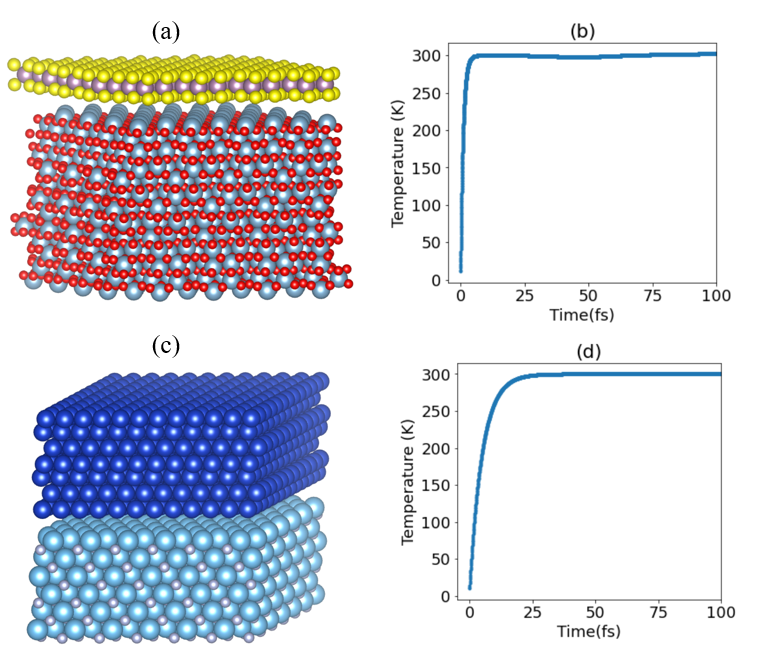}
    \caption{An example NVT simulation run for MoS$_2$(001)-Al$_2$O$_3$(001) and Cu(111)-TiN(111) interfaces. Fig. a and c show snapshots of the atomic structure and Fig. b and d show the time-evolution of temperature for the systems.}
\end{figure}

Next, we show example constant-temperature, constant-volume ensemble (NVT) simulation runs for MoS$_2$(001)-Al$_2$O$_3$(001) and Cu(111)-TiN(111) interface in Fig.6 using NVT ensemble and Berendsen thermostat. Several ensembles such as NPT, NVT and NVE with several structure optimizers are already available in ALIGNN-FF using the ASE package. The ALIGNN-FF is currently implemented as an ASE calculator which can be used for several tasks. In this particular MoS$_2$-Al$_2$O$_3$ example, we first generate the interfaces with JARVIS-Tools and then convert them into ASE-atoms object which is then subjected to structure optimizer and Berendsen thermostat at 300 K for 0.1 ps with 0.01 fs timestep. We observe how the thermostat gradually increases the temperature of the system to 300 K and then maintains the temperature, as expected. The structure keeps its integrity, \textit{i.e.,} doesn't collapse.

\subsection{Timing study}
Now, we compare the time for a single step potential energy calculation for FCC Aluminum (JVASP-816) with varied supercell sizes with EAM, ALIGNN-FF and GPAW DFT \cite{enkovaara2010electronic} methods. All of these calculations are performed on CPU on a personal laptop; GPU performance scaling may differ, particularly for large cell sizes. We start with a unit cell and start making supercells with [2,2,2], [3,3,3]. [4,4,4] and [5,5,5] dimensions. For GPAW case we use 8x8x8 k-points and as we make supercells, we reduce the k-points inversely proportional to the supercell size. In Fig. 7, we find that out of these three, EAM is the fastest method. The ALIGNN-FF is an order of magnitude slower than EAM, as expected. While EAM potentials are considerably faster, we note that they are difficult to train for multi-component systems. The GPAW single step calculations are almost 100 times slower than ALIGNN-FF.  As we add multiple electronic and ionic relaxation steps in DFT, the computational cost drastically increases. Note that we chose a very generalized DFT set up of 330 eV plane wave cutoff and PBE \cite{perdew1996generalized} exchange correlation functional for GPAW calculations. As we add more k-points and plane wave cut-off, the computational cost will increase.

 \begin{figure}[hbt!]
    \centering
    \includegraphics[trim={0. 0cm 0 0cm},clip,width=1.0\textwidth]{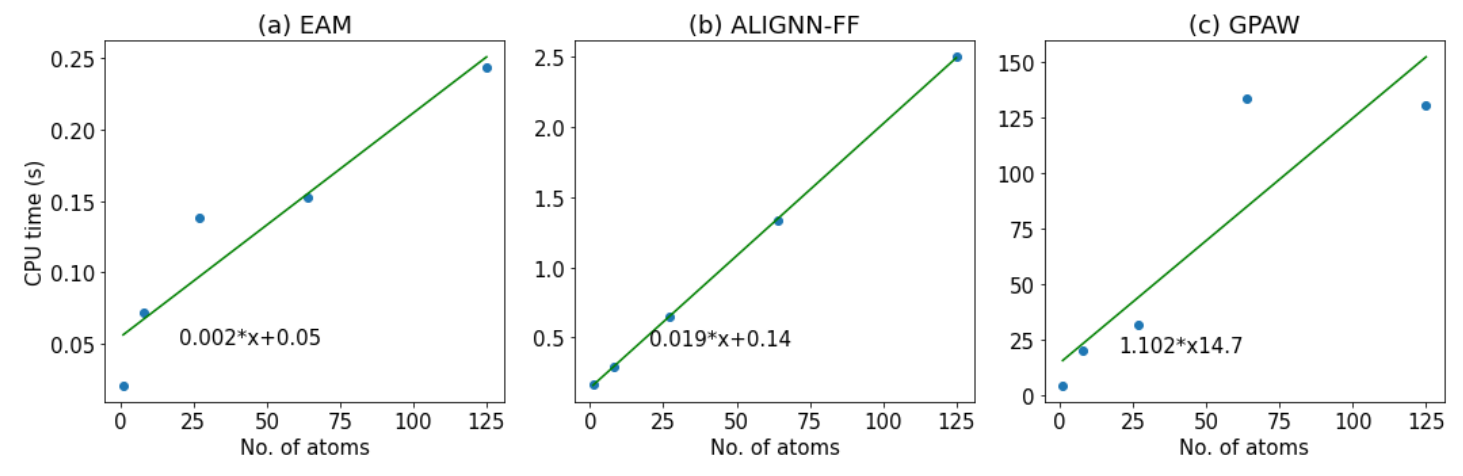}
    \caption{Timing comparison for FCC Al system for EAM, ALIGNN-FF and GPAW DFT methods.}
\end{figure}

\section{Conclusion}

In summary, we have developed a unified atomistic line graph neural network based force-field (ALIGNN-FF) that can model diverse set of materials with any combination of 89 elements from the periodic table. Using several test cases, we demonstrate the application of this method to determine the several properties such as lattice constants, formation energies, EV-curve for different materials including metallic, ionic and van der Waal bonded materials. 

Although, the above examples are based on a few test cases, ALIGNN-FF can in principle be used for several applications such as investigating defective systems, high-entropy alloy, metal-organic framework, catalyst, battery designs etc. and its validity needs to be tested for other applications which is beyond the scope of the present work. Also, as the methods for including larger datasets improve (such as training on millions of data), and integration of active learning and transfer learning strategies are achieved, we believe we can train more accurate models. Moreover, such universal FF model can be integrated with universal tight-binding model \cite{garrity2021fast} so that for classical and quantum properties can be predicted for large systems.
In the current state, such FFs can be very useful for structure optimization, however, there is lot of room for improvement in terms of other physical characteristics such as defects, magnetism, charges, electronic levels etc. Nevertheless, we believe such multi component interface simulations will greatly expedite high-throughput computational screening of advanced industrial materials.

% \subsection{Validation}

% \subsection{Predictive accuracy: Energy and Forces}
% \paragraph{E-V curve analysis}
% \paragraph{polymorph stability analysis}

% \paragraph{melting analysis}

% \subsection{Heterointerfaces}

% \section{Methods}
% \subsection{Dataset details}

% \subsection{ALIGNN model specification}

% \subsection{Training Details}

% \subsection{Evaluation Criteria}

%%%%%%%%%%%%%%%%%%%%%%%%%%%%%%%%%%%%%%%%%%%%%%%%%%%%%%%%%%%%%%%%%%%%%
%% The "Acknowledgement" section can be given in all manuscript
%% classes.  This should be given within the "acknowledgement"
%% environment, which will make the correct section or running title.
%%%%%%%%%%%%%%%%%%%%%%%%%%%%%%%%%%%%%%%%%%%%%%%%%%%%%%%%%%%%%%%%%%%%%
\begin{acknowledgement}

K.C. thanks National Institute of Standards and Technology for funding and computational support. In addition, K.C thanks Ruth Pachter and Kiet Ngyuen from the Air Force Research Laboratory for helpful discussions. This work was also partially supported by Wave 1 of the UKRI Strategic Priorities Fund under the EPSRC grant EP/T001569/1, particularly the ‘AI for Science’ theme within that grant, by the Alan Turing Institute and UKRI AIMLAC CDT, grant no. EP/S023992/1.

\end{acknowledgement}

%%%%%%%%%%%%%%%%%%%%%%%%%%%%%%%%%%%%%%%%%%%%%%%%%%%%%%%%%%%%%%%%%%%%%
%% The same is true for Supporting Information, which should use the
%% suppinfo environment.
%%%%%%%%%%%%%%%%%%%%%%%%%%%%%%%%%%%%%%%%%%%%%%%%%%%%%%%%%%%%%%%%%%%%%
% \begin{suppinfo}

% This will usually read something like: ``Experimental procedures and
% characterization data for all new compounds. The class will
% automatically add a sentence pointing to the information on-line:

% \end{suppinfo}

%%%%%%%%%%%%%%%%%%%%%%%%%%%%%%%%%%%%%%%%%%%%%%%%%%%%%%%%%%%%%%%%%%%%%
%% The appropriate \bibliography command should be placed here.
%% Notice that the class file automatically sets \bibliographystyle
%% and also names the section correctly.
%%%%%%%%%%%%%%%%%%%%%%%%%%%%%%%%%%%%%%%%%%%%%%%%%%%%%%%%%%%%%%%%%%%%%
\bibliography{alignn-ff}

\end{document}